\documentclass[number,3p,twocolumn]{elsarticle}

\usepackage{graphicx}

\usepackage{amssymb}
\usepackage{amsmath}
\usepackage{mathbbol}

\def\Tr{\mbox{Tr}}

\begin{document}

\begin{frontmatter}

\title{Thermodynamic anomalies in open quantum systems:\\ Strong coupling
effects in the isotropic XY model}


\author{Michele Campisi \fnref{label1}}
\author{ David Zueco   \fnref{label2}}
\author{Peter Talkner \fnref{label1}}

\address[label1]{Institut f\"ur Physik, Universit\"at Augsburg,
  Universit\"atsstra{\ss}e~1, D-86135 Augsburg, Germany}

\address[label2]{ Instituto de Ciencia de Materiales de Arag\'on y Departamento de F\'{\i}sica de la Materia Condensada, CSIC-Universidad de
  Zaragoza, E-50009 Zaragoza, Spain}

\begin{abstract}
The exactly solvable model of a one dimensional isotropic XY spin chain 
is employed to study the thermodynamics of open systems. For this purpose the chain is subdivided into two parts, one part 
is considered as {\it the system} while the rest as
{\it the environment} or {\it bath}.  The
equilibrium properties of the system display several anomalous aspects such as negative entropies, negative
specific heat, negative susceptibilities in dependence of temperature  and coupling strength between system and bath.
The statistical mechanics of this system is studied
in terms of a reduced density matrix. At zero temperature and for a certain parameter values we observe a change of the ground state, a situation akin to a quantum phase transition.
\end{abstract}

\end{frontmatter}

\section{Introduction}
\label{sec:intro}

Ordinary thermodynamics describing the macroscopic phenomenology of homogeneous equilibrium systems leads to the following \emph{stability conditions}:
\begin{eqnarray}
\partial P/\partial V & \leq & 0 \label{eq:dP/dV<0}\\
C_V &\geq& 0 \label{eq:CV>0}
\end{eqnarray}
These conditions ensure that the pressure $P$ of a system decreases as its volume $V$ increases and that the system warms up when absorbing energy at fixed volume (positive specific heat at constant volume $C_V$).
One of the major achievements of mathematical physics was to show that ordinary matter, composed of a large number of electrons and protons, behave according to Eqs. (\ref{eq:dP/dV<0},\ref{eq:CV>0}) \cite{LebowitzLiebPRL69}.

However, it has recently turned out that systems not satisfying the condition (\ref{eq:CV>0}), actually exist in nature, and that a corresponding thermodynamic description should yet be possible in certain cases \cite{HIT_NJP08,IHT_PRE09}.
The prototypical examples of such unstable systems are stars which are known to expand and cool down as their energy increases \cite{Lynden-Bell99}.

Within the canonical ensemble the specific heat is related to the fluctuations of energy via the relation $k_B C_V= T^{-2}\langle \delta E^2 \rangle$. From this expression it is evident that specific heat is necessarily positive in the canonical ensemble. Negative specific heat, however may appear within the microcanonical ensemble. This can happen due to different mechanisms:


\begin{itemize}
 \item The ergodic properties of the system may depend on energy. Hence, at different energies, different parts of phase space are accessible. This may lead to negative specific heat and other thermodynamical anomalies.
An example was given by Hertel and Thirring \cite{HertelThirring71}.

\item
The system might be far from the thermodynamic limit. Systems that are stable in the thermodynamic limit, such as Lennard-Jones gases, may display negative specific if only their size is small enough \cite{Thirring03,DunkelPRE06}.

\item Long ranged forces might prevent the thermodynamic limit to exist at all. But even when an equilibrium state exists such systems remain \emph{nonextensive} and may show negative specific heat \cite{Ruffo_PRL01,Borges_PHYSA}.

\end{itemize}

In this work we will present yet another mechanism that leads to negative specific heat and other thermodynamic anomalies for systems in {\it contact} with a {\it heat bath}.
Apart from extensivity and short ranged interactions, another assumption is customarily made in the statistical mechanics of canonical systems, namely \emph{weak coupling} between system and its environment. When the coupling to the environment is not negligible, violations of condition (\ref{eq:CV>0}) may appear even if system and environment, as a whole, are in a canonical state \cite{HIT_NJP08,IHT_PRE09}. Hence, yet another item can be added to our list of exceptions:

\begin{itemize}
 \item  Systems that strongly interact with their environment may display thermodynamic anomalies
\end{itemize}

For example, a single free particle, which would display a positive specific heat when weakly coupled to a bath, may display negative specific heat when in strong interaction with the bath \cite{HIT_NJP08}.
Similar effects were observed for a two level system coupled to  a harmonic oscillator \cite{CTH_JPHYSA09}.

We assume that the total system $\mathcal{S}+\mathcal{B}$, is in \emph{weak} contact with a \emph{super-bath} that provides the temperature concept ($T= (k_B \beta)^{-1}$, $k_B$ is Boltzmann constant.)
This total system is therefore described by the \emph{canonical} statistics $e^{-\beta H_{tot}}/Z_{tot}$, where $H_{tot}$ is the Hamiltonian of the total system $\mathcal{S}+\mathcal{B}$ and $Z_{tot}=\Tr e^{-\beta H_{tot}}$. Accordingly, its specific heat cannot be negative.
However, if the coupling between $\mathcal{S}$ and $\mathcal{B}$
is non-negligible, the overall canonical state does not factorize into the product of two canonical states for $\mathcal{S}$ and $\mathcal{B}$ respectively: $e^{-\beta H_{tot}}/Z_{tot} \neq e^{-\beta H_{S}}e^{-\beta H_{B}}/(Z_{S}Z_{B})$. 
Although the system is in perfect thermal equilibrium with its environment, it is not in a canonical state and therefore negative entropies and violations of the inequalities (\ref{eq:dP/dV<0},\ref{eq:CV>0}) may occur.
These violations though are not a sign of any instability because as a part of a stable total system, the system of interest itself is also stable.

\begin{figure}[t]
\includegraphics[width=7.5cm]{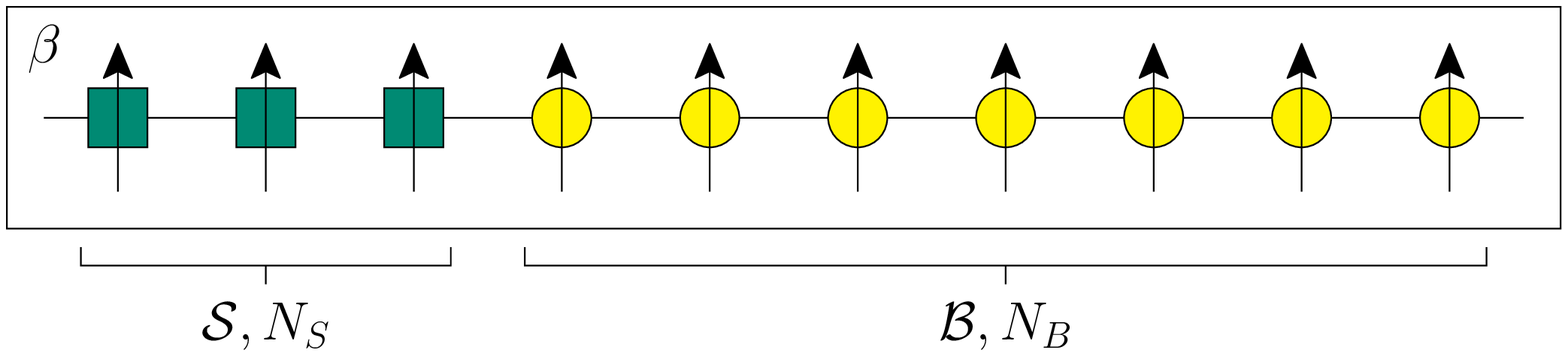}
\caption{(Color online) Schematic representation of the model studied. A spin chain is immersed in a thermal environment (referred to as the {\it super-bath} in the text) at temperature $T=1/(k_B \beta)$. The chain is composed of two parts: The system of interest $\mathcal{S}$, made of the first left-most $N_S$ spins (green squares), and the ``bath'' $\mathcal{B}$, made of the remaining $N_B$ spins (yellow circles).
When the interaction between the two subchains is non negligible, anomalies may occur in the thermodynamics of the system of interest $\mathcal{S}$.}
\label{fig:lattice}
\end{figure}

There are few available exact solutions of open systems displaying this kind of anomalies. One known example is the damped free particle \cite{HIT_NJP08, IHT_PRE09,HI_ActPhysPol06} and a two-level fluctuator in contact with a single oscillator \cite{CTH_JPHYSA09}. Numerical investigations of thermodynamic anomalies in the context of the Casimir effect, the multichannel Kondo effect and of mesoscopic superconductors containing magnetic impurities have been reported recently in the literature \cite{Hoye_PRE03,Florens_PRL04,Zitko_PRB09}.
In this work we study the thermodynamics of a subchain of a longer chain of spins interacting according to the isotropic XY
model with free ends \cite{LiebShultzMattis61, mikeska}. 
The spin chain is composed of two parts. One subchain is defining the system
$\mathcal S$, while the rest of the chain comprises the bath, $\mathcal
B$ (see Fig. \ref{fig:lattice}).  Being this an exactly solvable model, we are able to analytically find
the relevant thermodynamical functions for the subchain.

Far from being a  purely academic problem, the study of the equilibrium properties of spin-$1/2$ chains has been recently attracting a great deal of attention. Spin systems not only are the basis of the physics of magnetic materials \cite{White1983a} but they
might have an enormous impact with regard to the development of quantum technologies \cite{Amico2008, Latorre, Zueco_PRE09,Galve_PRA09,Saito_PRB07}.
We believe that the study of the thermodynamics of small quantum systems will help to understand problems of quantum information and vice versa
\cite{Acin}.

In Sec. \ref{sec:TDopen} we briefly review the generalities of the thermodynamics of open systems. The specific model studied in this paper is described in Sec. \ref{sec:model}, while its thermodynamics is illustrated in Sec. \ref{sec:thermodynamics}. Various anomalies ranging from negative entropy and negative specific heat, to negative susceptibility are observed for system size, $N_S$, of the order of unity. In Sec. \ref{sec:rho} we study the reduced density matrix of the open system $\mathcal S$ and show that it departs from the canonical form in the strong coupling regime. The spectrum of the reduced density matrix is analyzed and it is observed that, regardless of the strong coupling, a quantum phase transition occurs at zero temperature.
Conclusions will be drawn in Sec. \ref{sec:conclusions}.

\section{Thermodynamics of open systems}
\label{sec:TDopen}
Consider the following Hamiltonian:
\begin{equation}
\label{Htotal}
 H_{tot}= H_S +H_B +H_{SB}
\end{equation}
describing a systems of interest $\mathcal S$ interacting with a second system, the bath $\mathcal B$, via the interaction energy term $H_{SB}$.
The compound system $\mathcal{S}+\mathcal{B}$ is in weak contact with a large super-bath at temperature $T=1/{k_B \beta}$. Hence it is described by the canonical density matrix:
\begin{equation}
\rho_{tot} = e^{-\beta H_{tot}}/Z_{tot}
\end{equation}
where 
\begin{equation}
Z_{tot}= \Tr \, e^{-\beta H_{tot}}
\end{equation}
is the total system partition function, with $\Tr$ denoting the trace over the total system Hilbert space.
The partition function of the open quantum system, $\mathcal{S}$, is given by the ratio of the total system partition function $Z_{tot}$ and the bare bath partition function $Z_B$, i.e.
\cite{HIT_NJP08,IHT_PRE09,CTH_JPHYSA09,Campisi2009a,HaenggiQTransp98,
Feynman:1963qm,Ford:1985it,Caldeira:1983aj}:
\begin{equation}
Z = {Z_{tot}}/{Z_B}
\label{eq:Z=Ztot/ZB}
\end{equation}
where 
\begin{equation}
Z_{B}= \Tr_B \,e^{-\beta H_{B}}
\end{equation}
with $\Tr_B$ denoting the trace over the bath Hilbert space.
According to the rules of statistical mechanics, the Helmholtz free energy of the open quantum system is then:
\begin{equation}
F= -\beta^{-1} \ln Z \,.
\label{eq:F}
\end{equation}
From this free energy, all the relevant thermodynamic functions of the open quantum system can be derived.
In particular, the entropy and the specific heat are given by $S=-\partial F/ \partial T$, and $C_V=T \partial S/\partial T$, respectively.
Thanks to Eq. (\ref{eq:Z=Ztot/ZB}) the system's free energies is the difference of the free energy of the total system and the bare bath, i.e.:
\begin{equation}
F= F_{tot}-F_{B} \, .
\label{eq:F=Ftot-FB}
\end{equation}
Due to the linearity of the derivative, this implies according relations for the entropy and the specific heat of the system, which take the form: 
\begin{align}
S &= S_{tot}-S_{B} \, ,\\
C &= C_{tot}-C_{B}\, .
\end{align}
While $C_{tot}$ and $C_B$ are positive numbers, there is no reason 
why their difference should be positive as well. Hence, coupling the bath $\mathcal B$ strongly to the system $\mathcal S$ may result in an overall decrease of specific heat. The same holds for the entropy and other thermodynamic quantities that linearly depend on the free energy. These relations will be exemplified with the model described below.

\section{The Model}
\label{sec:model}
We consider the isotropic $XY$ model of $N$ interacting spins on a one
dimensional lattice of equally spaced sites with {\it free ends}:
\begin{equation}
\label{XYmodel}
H_N = \frac{h}{2} \sum_{j=1}^{N} \sigma^z_j +\frac{J}{2} \sum_{j=1}^{N-1}
(\sigma^x_j \sigma^x_{j+1}+\sigma^y_j \sigma^y_{j+1})
\end{equation}
where $\sigma^x_j,\sigma^y_j$ and $\sigma^z_j$ are the Pauli matrices of
the $j^{th}$ spin. 
Dropping the label
$j$ these are:
\begin{eqnarray}
\sigma^x &=& \left(\begin{array}{cc}
0 & 1\\
1 & 0 \end{array}\right) \, ,\\
\sigma^y &=& \left(\begin{array}{cc}
0 & -i\\
i & 0 \end{array}\right)\, ,\\
\sigma^z  &=& \left(\begin{array}{cc}
1 & 0\\
0 & -1 \end{array}\right)\, .
\end{eqnarray}
The first term in the Hamiltonian (\ref{XYmodel}) accounts for the
{\it Zeeman} interaction between each spin and an
applied magnetic field $h$ pointing in the $z$-direction.
The second term describes the nearest neighbor interaction with
strength $J$.

The isotropic $XY$ model (\ref{XYmodel}) is an exactly solvable model. By means of the Jordan-Wigner transformation \cite{mikeska}, the Hamiltonian (\ref{XYmodel}) is mapped onto a system of $N$ free fermionic eigenmodes, of energies
\begin{equation}
\lambda_k^{(N)} = h - 2J \cos \left( \frac{k\pi}{N+1} \right) \qquad k=1 \dots N\, ,
\label{eq:lambda}
\end{equation}
see \ref{App:JW}. Thereby, the spectrum of the Hamiltonian (\ref{XYmodel}) is expressed in terms of the occupation numbers of each fermionic mode $n_j$, as
\begin{equation}
\label{spectrum}
\varepsilon_{n_1, \dots n_N} = \sum_{j=1}^{N}\lambda_k^{(N)}  n_j + \frac{N h}{2}
\end{equation}
where each $n_j$ can only be $0$ or $1$.
Given the spectrum, the partition function, $Z_N = \sum_{n_1, \dots n_N}e^{-\beta
\varepsilon_{n_1, \dots n_N}}$, is calculated as:
\begin{equation}
Z_N = e^{-\beta N h/2}\prod_{k=1}^{N} \left( 1 + e^{-\beta
\lambda_k^{(N)}}\right)\, .
\end{equation}
According to the formula $F=-\beta^{-1}\ln Z$, the free energy of an isotropic $XY$ chain of $N$ spins then becomes:
\begin{equation}
F_N =\sum_{k=1}^{N} f_k^{(N)}
\label{eq:FN}
\end{equation}
where
\begin{equation}
f_k^{(N)}= -\beta^{-1}\ln \left( 1 + e^{-\beta
\lambda_k^{(N)}}\right)+ \frac{h}{2}
\label{eq:fk}
\end{equation}
denotes the contribution from the $k^{\text{th}}$ fermionic mode.

\subsection{A part of the chain as an open system}

We now consider the part of the chain consisting of the first $N_S$ spins, counted, say, from the left end of the chain, as the system of interest, $\mathcal{S}$, and the rest of $N-N_S$ spins as the bath, $\mathcal B$, see Fig. \ref{fig:lattice}.
The total Hamiltonian $H_N$, can be recast in the usual system-bath form $H_N=H_S+H_B+H_{SB}$ of Eq. (\ref{Htotal}), with
\begin{equation}
\label{Hs}
H_S = \frac{h}{2} \sum_{j=1}^{N_S} \sigma^z_j -\frac{J}{2} \sum_{j=1}^{N_S-1}
(\sigma^x_j \sigma^x_{j+1}+\sigma^y_j \sigma^y_{j+1})\, ,
\end{equation}
\begin{equation}
\label{Hb}
H_B = \frac{h}{2} \sum_{j=N_S+1}^{N_S} \sigma^z_j -\frac{J}{2}
\sum_{j=N_S+1}^{N-1} (\sigma^x_j \sigma^x_{j+1}+\sigma^y_j
\sigma^y_{j+1})\, ,
\end{equation}
\begin{equation}
\label{HSB}
H_{SB} = -\frac{J}{2} (\sigma^x_{N_S} \sigma^x_{N_S+1}+\sigma^y_{N_S}
\sigma^y_{N_S+1})\, .
\end{equation}
The regime of strong coupling holds when the system energy is comparable to the interaction energy. Weak coupling is achieved in this problem when at least one of the following holds $N_S \gg 1$, $h \gg J$, $k_BT \gg \max(h,J)$.


From the Eq. (\ref{eq:F=Ftot-FB}), the thermodynamic free
energy of the open system $\mathcal{S}$ is the difference between the free
energy of the total chain of length $N$ and the free energy of the
bare bath $\mathcal{B}$, which is itself a $XY$ chain of a certain
length $N_B$. Its free energy is then readily obtained by replacing
$N$ with $N_B$ in Eq. (\ref{eq:FN}). We thus obtain for the system of
interest $\mathcal{S}$:
\begin{equation}
 F= F_N-F_{N_B}\, .
\end{equation}
Using Eqs. (\ref{eq:FN},\ref{eq:fk}):
\begin{equation}
 F= \sum_{k=1}^{N} f_k^{(N)} -\sum_{q=1}^{N_B} f_q^{(N_B)}\, .
\end{equation}

\section{Thermodynamics of the subchain $\mathcal{S}$}
\label{sec:thermodynamics}
We discuss now the explicit results for the relevant
thermodynamical functions, such as entropy, specific heat, magnetization and susceptibility of the open system defined in
Eqs. (\ref{Htotal}) and (\ref{Hs}).  See also Fig. \ref{fig:lattice}.
\subsection{Entropy}
\label{subsec:entropy}
We begin our discussion with studying the entropy of $\mathcal{S}$:
\begin{equation}
S = -\left. \frac{\partial F}{\partial T}\right|_h \, .
\label{eq:S}
\end{equation}
From Eq. (\ref{eq:FN}), follows
\begin{equation}
S = S_N-S_{N_B}=  \sum_{k=1}^{N} s_k^{(N)} -\sum_{q=1}^{N_B} s_q^{(N_B)}
\label{eq:S_2}
\end{equation}
where $S_N$ and $S_{N_B}$ denote the total system and the bare bath entropies, respectively, and 
\begin{equation}
 s_k^{(N)}=\ln \left(1+e^{\beta \lambda_k^{(N)}}\right)+
\frac{\beta \lambda_k^{(N)}}{1+e^{\beta \lambda_k^{(N)}}}
\label{eq:sk}
\end{equation}
is the single mode entropy of the total chain of length $N$. Likewise $s_q^{(N_B)}$ is the single mode entropy of the bath $\mathcal{B}$.
\begin{figure}[t]
\includegraphics[width=8cm]{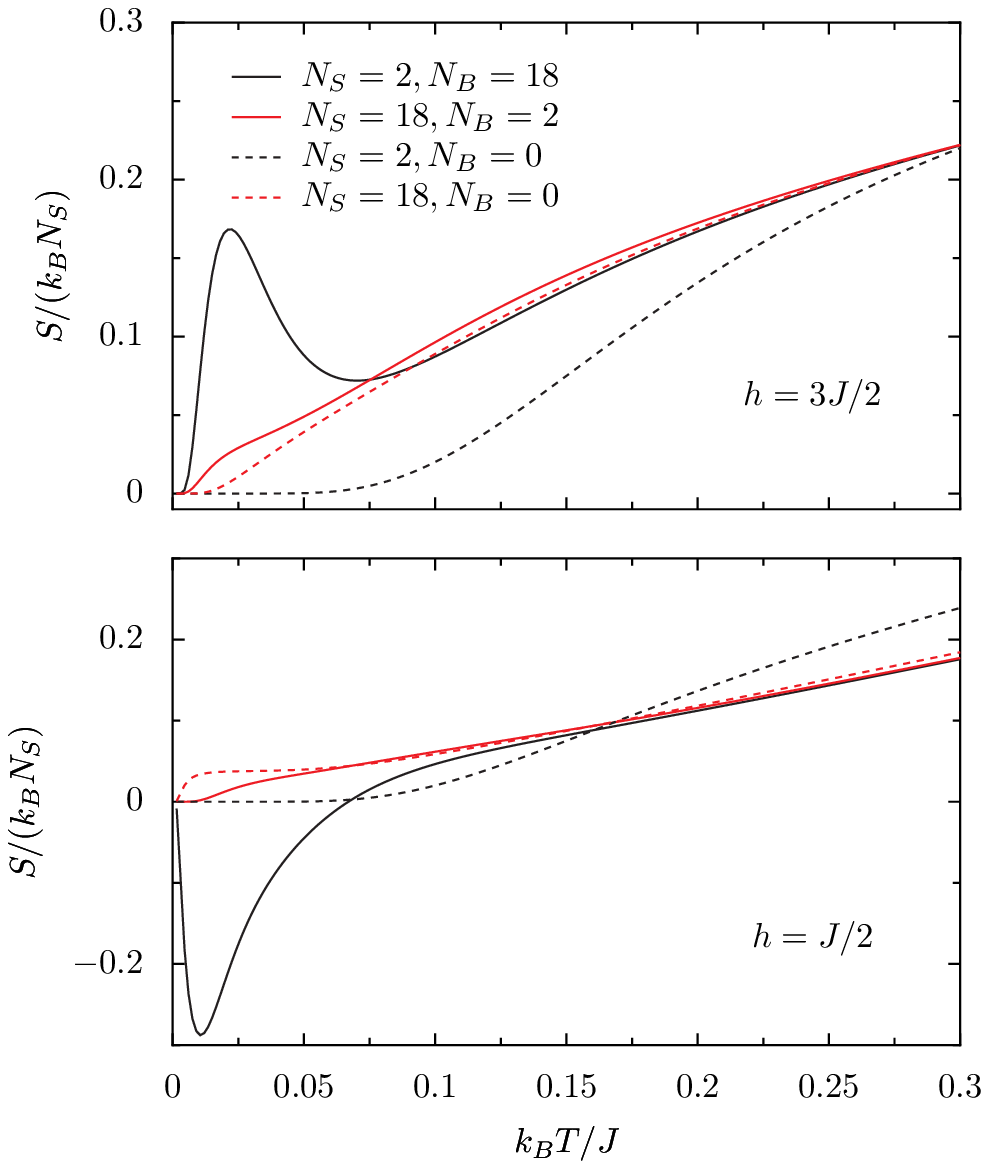}
\caption{(Color online) Entropy per spin for $N_S=2,N_B=18$ (black solid line), $N_S=18,N_B=2$ (red solid line), $N_S=2,N_B=0$ (black dashed line), $N_S=18,N_B=0$ (red dashed line). Top panel: $h=3J/2$, bottom panel $h=J/2$. Entropy becomes negative at low temperature for the lower magnetic field $h=J/2$, in the strongly coupled case, i.e. $N_S=2,N_B=18, h=J/2$}
\label{fig:S}
\end{figure}
In Fig. \ref{fig:S} the temperature dependence of the entropy per spin
$S/N_S$ of the system $\mathcal{S}$, is shown for different system sizes $N_S$ of
$\mathcal{S}$ and fixed total size $N$. The magnetic field is $h=3J/2$  and  $h=J/2$ in the top and bottom panel of Fig. \ref{fig:S}, respectively.
For comparison, the graphs showing the temperature dependence of entropy for the
same system sizes in absence of the bath, are shown as well. We shall use the label $\mathcal{S}_0$ for the bare system.

Notably the presence of the bath does not affect much the entropy of the system
at high temperatures, whereas its effect is most prominent at low temperatures.
This high temperature behavior can be understood by looking at Eq. (\ref{eq:S_2}).
The entropy $s_k^{(N)}$ of each mode tends to $\ln 2$ for large temperatures, thus the
entropy of the total system tends to $N\ln 2$. Likewise, the entropy of the bare bath tends to 
$N_B \ln 2$ and that of the system $\mathcal{S}$ tends to $(N-N_B) \ln 2$, which
is the same as the entropy of the bare system $\mathcal{S}_0$ at high temperature, i.e., $N_S \ln 2$. 
Thus, for increasing temperature the effect of the bath on the entropy of the
system $\mathcal{S}$ becomes less relevant.

With reference to both panels of Fig. \ref{fig:S}, we see that at low temperature, the effect of the bath becomes
increasingly relevant as the system size decreases.
As the system shrinks, the relative effect of the interaction with the bath,
which is a \emph{surface} effect, becomes more important (strong
coupling).
In particular, we notice the pronounced peak in the entropy
for the smallest value of $N_S$ ($N_S=2$) in  the case $h=3J/2$, top panel of Fig. \ref{fig:S}.
A different situation occurs in the bottom panel of Fig. \ref{fig:S} where the magnetic
field is chosen to be $h=J/2$ and all other parameters are kept unchanged as
compared to Fig. \ref{fig:S}.
The major difference is the presence of a region of \emph{negative
entropy}.
This is a feature that is often found in open systems, in the regime of strong
coupling \cite{HIT_NJP08,IHT_PRE09,Hanggi2005b,CTH_JPHYSA09}
\begin{figure}[t]
\includegraphics[width=8cm]{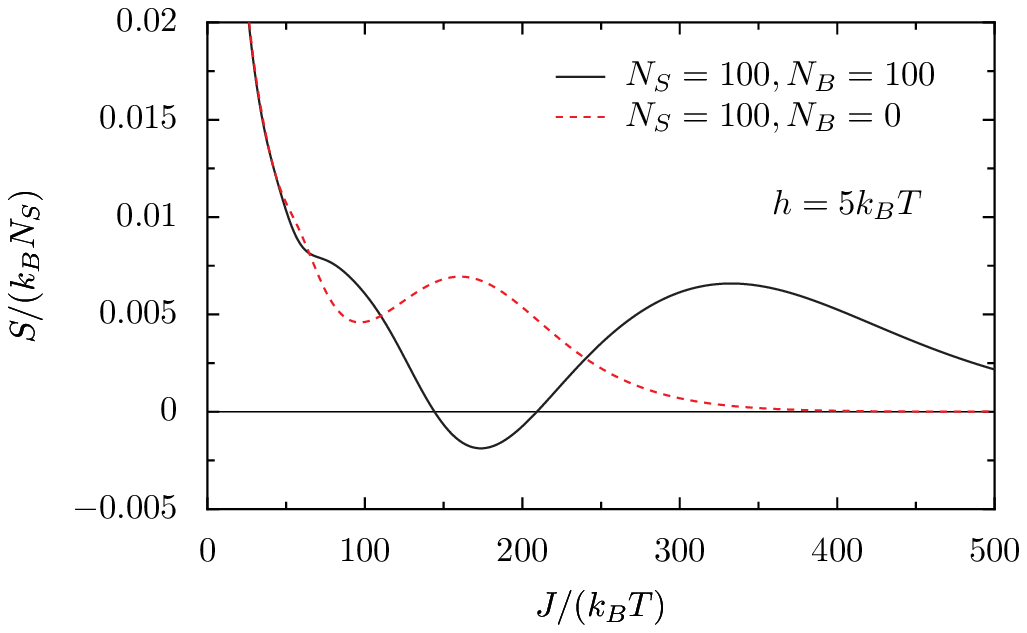}
\caption{(Color online) Entropy per spin as a function of $J/(k_B T)$ for chain of  $N_S=100$ spins coupled to a bath of $N_B=100$ spins (black solid line), as compared to an isolated chain with  $N_S=100$ (red dashed line) at fixed $h=5 k_B T$. The two entropies coincide only for small enough values of $J/(k_B T)$. While the entropy of the uncoupled chain is always positive, the entropy of the coupled chain may become negative for large enough values of $J/(k_B T)$.}
\label{fig:SJ}
\end{figure}

Fig. \ref{fig:SJ} shows the entropy of a larger chain ($N_S=100$) coupled to an equally large bath ($N_B=100$) as a function of coupling strength $J/(k_B T)$, as compared to the entropy of the same chain with no bath. For small enough $J/(k_B T)$ the two entropies are approximately equal, while thir difference is apparent for large enough coupling. The entropy of the uncoupled chain is always positive, whereas the entropy of the coupled chain may become negative for large enough coupling $J/(k_B T)$.

\subsection{Specific heat}
\label{subsec:specificheat}
The specific heat at constant magnetic field $h$ is obtained from the entropy
via the standard formula:
 \begin{equation}
C = T \left.\frac{\partial S}{\partial T}\right|_h.
\end{equation}
Evidently $C$ is the difference between the total system specific heat, $C_N$,  and the bare bath specific heat, $C_{N_B}$
\begin{equation}
C = C_N-C_{N_B}=  \sum_{k=1}^{N} c_k^{(N)} -\sum_{q=1}^{N_B} c_q^{(N_B)}
\label{eq:C_2}
\end{equation}
where the single mode specific heat is:
 \begin{equation}
c_{k}^{(N)}= T\left.\frac{\partial  s_k^{(N)}}{\partial T}\right|_h=
\left(
\frac{\beta \lambda_k^{(N)}/2 }{\cosh(\beta\lambda_k^{(N)}/2)}
\right)^{2}
\end{equation}
The specific heat of each single mode vanishes at high temperature and so do the
total chain specific heat, the specific heat of the bath, the specific heat of
the system $\mathcal{S}$, and the specific heat of the bare system
$\mathcal{S}_0$.

\begin{figure}[t]
\includegraphics[width=8cm]{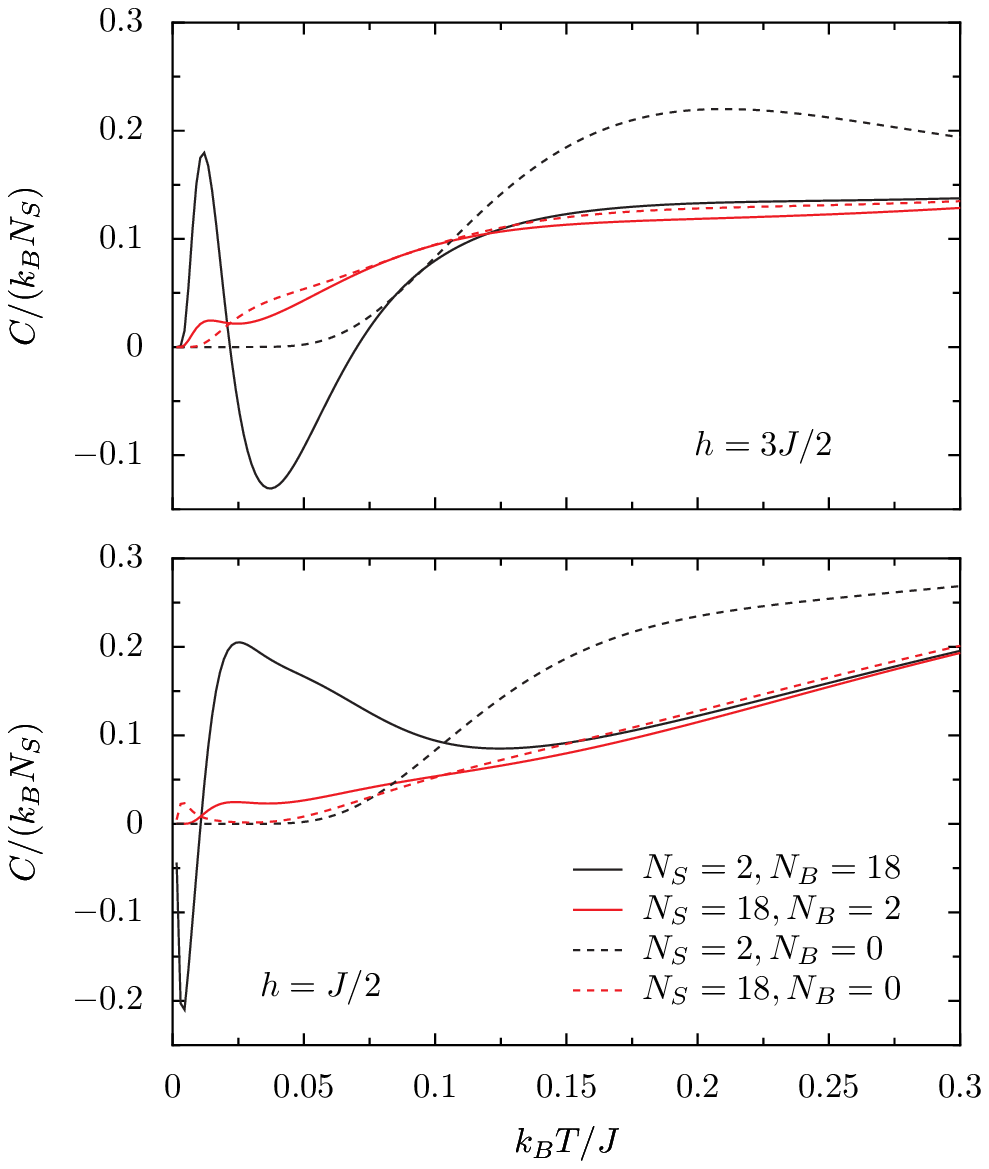}
\caption{(Color online) Specific heat per spin for $N_S=2,N_B=18$ (black solid line), $N_S=18,N_B=2$ (red solid line), $N_S=2,N_B=0$ (black dashed line), $N_S=18,N_B=0$ (red dashed line). Top panel: $h=3J/2$, bottom panel $h=J/2$. Negative specific heat regions appear at low temperature in the case $h=J/2$ (bottom) and at intermediate temperatures in the case $h=3J/2$.}
\label{fig:C}
\end{figure}

Fig. \ref{fig:C} shows the specific heat for the same parameters as in Fig.
\ref{fig:S}, that is $N=20$, $N_S=2,18$, with $h=3J/2$ (top panel) and $J/2$ (bottom panel).
Quite interestingly, for the smallest system ($N_S=2$) which is more affected by the bath, the specific heat displays a very pronounced positive peak at very low
temperature followed by a \emph{negative} peak at intermediate temperatures, for the case $h=3J/2$.
This appearance of a negative specific heat at intermediate temperature was
observed also for a free particle coupled to a minimal bath composed of a single
oscillator \cite{HIT_NJP08}.

The behavior changes if the magnetic field is decreased to a value smaller than
$J$, e.g. $h=J/2$, see bottom panel in Fig. \ref{fig:C}.
In case of very small $N_S$, i.e. $N_S=2$, there is a negative peak at very low
temperature and a positive one at intermediate temperature.
A similar situation was also observed for the case of a two-level system coupled
to a minimal bath composed of a single oscillator \cite{CTH_JPHYSA09}.

The sign of the specific heat is given by the sign of the derivative of $S$ with
respect to temperature $T$. Thus, whenever the entropy displays a region where it
decreases for increasing temperature, correspondingly the specific heat displays
a region of negative values. Therefore, the presence of a positive peak in the entropy
at low temperature leads to the appearance of negative and positive peaks
in the specific heat at low and intermediate $T$'s, respectively (compare the top panels in Figs. \ref{fig:S}, \ref{fig:C}). Likewise the presence of a negative peak in the entropy at
low $T$ leads to a negative specific heat region at low temperature whose
extension is smaller than that of negative entropy (see also \cite{CTH_JPHYSA09}).

\subsection{Magnetization}
\label{subsec:magnetization}

Of further interest is the magnetization of the system $\mathcal{S}$ as a function of
temperature $T$ and magnetic field $h$:
\begin{equation}
 M=-\left.\frac{\partial F}{\partial h}\right|_T \,.
\end{equation}
It is given by the difference of the magnetizations of the total chain and the part of the chain representing the bath, reading:
\begin{equation}
M = M_N-M_{N_B}=  \sum_{k=1}^{N} m_k^{(N)} -\sum_{q=1}^{N_B} m_q^{(N_B)} \, ,
\label{eq:M_2}
\end{equation}
where the single mode magnetization, $m_k^{(N)}$ is:
\begin{equation}
m_k^{(N)} = \frac{1}{2} \tanh\left({\frac{\beta \lambda_k^{(N)}}{2}}\right) \, .
\end{equation}
\begin{figure}[t]
\includegraphics[width=8cm]{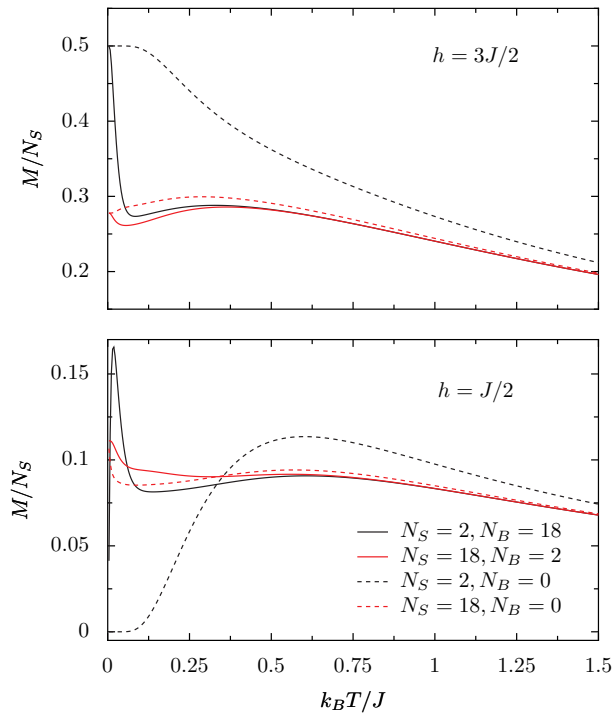}
\caption{(Color online) Magnetization per spin for $N_S=2,N_B=18$ (black solid line), $N_S=18,N_B=2$ (red solid line), $N_S=2,N_B=0$ (black dashed line), $N_S=18,N_B=0$ (red dashed line). Top panel: $h=3J/2$, bottom panel $h=J/2$.}
\label{fig:M}
\end{figure}
At high temperature the single mode magnetization tends to zero as the thermal
agitation wins over magnetic ordering.
Fig. \ref{fig:M} shows the magnetization per spin for the same parameters
reported in Figs. \ref{fig:S}, \ref{fig:C}, i.e., $N=20$, $N_S=2,18$ and
$h=3J/2$ (top panel), $h=J/2$ (bottom panel). 

With reference to the top panel, we see that in absence of a bath the magnetization per spin of the shortest
chain, $N_S=2$, is much larger than the magnetization of the longer chain
$N_S=18$, for thermal energies ($k_B T$) up to the order of $J$.
When the system is put in contact with the bath, no relevant
change in the magnetization is observed for the longer chain $N_S=18$.
For the smallest chain however, a quick drop of the magnetization occurs with increasing temperature, and values close to those pertaining to larger chains are reached already at temperatures of the order $J/(10 k_B)$.

The bottom panel of Fig. \ref{fig:M} shows the magnetization per spin for the same parameters as of the top panel but for
$h=J/2$. In contrast to the case $h=3J/2$, and with reference to the smallest
chain, $N_S=2$, we observe a dramatic enhancement of several orders of magnitude of magnetization at values of $k_B T/J \lesssim 0.1$, due to the coupling to the bath.

\subsection{Susceptibility}
\label{subsec:susc}

By taking the partial derivative of the magnetization $M$ with respect to
magnetic field, one obtains the magnetic susceptibility:
\begin{equation}
 \mathcal{X}=\left. \frac{\partial M}{\partial h}\right|_T \, .
\end{equation}
One finds:
\begin{equation}
\mathcal{X} = \mathcal{X}_N-\mathcal{X}_{N_B}=  \sum_{k=1}^{N} \chi_k^{(N)} -\sum_{q=1}^{N_B} \chi_q^{(N_B)} \, ,
\label{eq:x_2}
\end{equation}
where the single mode susceptibility reads:
\begin{equation}
\chi_k^{(N)}= \frac{\beta}{4 \cosh^2 (\beta \lambda_k^{(N)}/2)} \, .
\end{equation}

\begin{figure}[t]
\includegraphics[width=8cm]{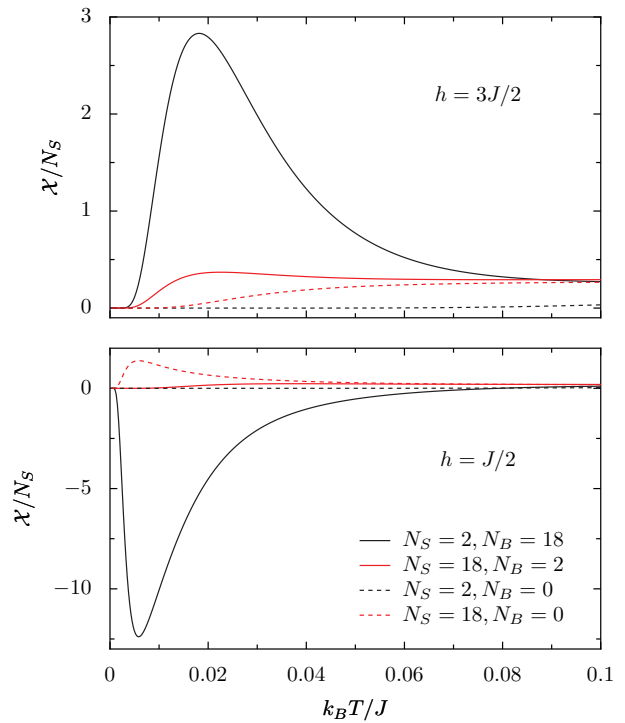}
\caption{(Color online) Susceptibility per spin for $N_S=2,N_B=18$ (black solid line), $N_S=18,N_B=2$ (red solid line), $N_S=2,N_B=0$ (black dashed line), $N_S=18,N_B=0$ (red dashed line). Top panel: $h=3J/2$, bottom panel $h=J/2$.}
\label{fig:X}
\end{figure}

Fig. \ref{fig:X} displays the susceptibility for the same set of parameter
values used in Figs. \ref{fig:S}, \ref{fig:C}, \ref{fig:M}, i.e.,
$N=20$, $N_S=2,18$ and
$h=3J/2, h=J/2$. Note the enhancement of susceptibility at low temperature in both cases due to the presence of the bath. As expected, this is more pronounced for the shortest chain.

Of particular interest is the behavior of the susceptibility for values of $h<J$
and the smallest chain (bottom panel of Fig. \ref{fig:X}). Note
the large \emph{negative} enhancement at low temperature for $N_S=2$.
Not only the entropy and the specific heat may display anomalous behavior in
open systems and take on negative values, but also the susceptibility may do so.
\subsection{Internal energy}
\label{subsec:U}
The system internal energy is obtained from the thermodynamic relation:
\begin{equation}
 U = F + T S
\end{equation}
Evidently, also the internal energy of the system is given by the difference of the respective quantities for the total system and the bare bath, resulting to:
\begin{equation}
U = U_N-U_{N_B}=\sum_{k=1}^{N} u_k^{(N)} -\sum_{q=1}^{N_B} u_q^{(N_B)}
\label{eq:U_2}
\end{equation}
where the single mode energy is given by:
\begin{equation}
u_k^{(N)} = f_k^{(N)} + T s_k^{(N)}\, .
\end{equation}
Using Eqs. (\ref{eq:fk},\ref{eq:sk}) for $f_k$ and $s_k$ we obtain
\begin{equation}
u_k^{(N)} = \mathcal{N}_k^{(N)} \lambda_k^{(N)} +h/2
\end{equation}
where
\begin{equation}
\mathcal{N}_k^{(N)} = \frac{1}{1+ e^{\beta \lambda_k^{(N)}}}
\label{eq:Nk}
\end{equation}
is the Fermi distribution giving the average occupation number of the $k$'th fermionic eigenmode.

\begin{figure}[t]
\includegraphics[width=8cm]{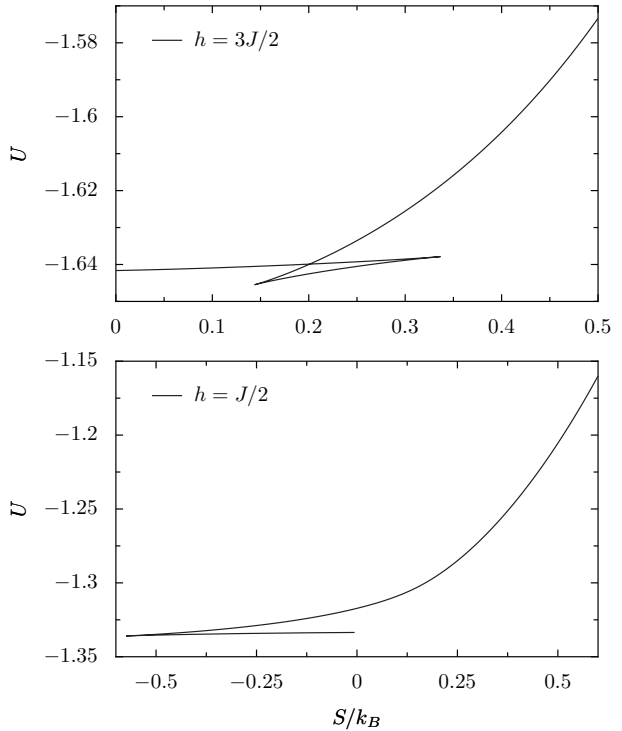}
\caption{Parametric plot of entropy versus internal energy. Cusps appear both at $h=J/2$ (bottom), and $h=3J/2$ (top). For both values of $h=J/2$ (bottom) and $h=3J/2$ (top) cusps appear where the entropy has a maximum as a function of temperature.}
\label{fig:U}
\end{figure}

In Fig. \ref{fig:U} we show the curve $(U(T,h),S(T,h))$,
parameterized by the temperature $T$, for the fixed parameters $N=20$, $N_S=2$ and
$h=3J/2$ (top), $h=J/2$ (bottom).
Note that the relation of entropy and internal energy is not one-to one everywhere, but there is a region of internal energies belonging to different entropies and vice-versa.
This is a specific feature of strongly coupled open systems, which may not
appear in the weak coupling regime.
The graph  $(U(T,h),S(T,h))$ contains many informations. 
Each point on the plotted curves corresponds to a given temperature $T$, which is
identical to the slope of the graph at the very same point.
\begin{equation}
 \frac{\partial U }{\partial S} = \frac{\partial U/\partial T}{\partial S /
\partial T}=\frac{\partial (F+TS)/\partial T}{\partial S / \partial T}=T
\label{Eq:dU/dS=T}
\end{equation}
Cusps appear in the curve $(U(T,h),S(T,h))$ for those values of $T$ for which
the entropy $S(T,h)$ has a local extremum. 
By comparison with Fig.
\ref{fig:S},
we see that for $N_S=2$, $h=3J/2$, first the
entropy goes through a maximum and subsequently a minimum, as the temperature is increased. These two extrema
correspond to the two cusps one encounters in the graph $(U(T,h),S(T,h))$ as one follows the curve in the $U$-$S$ plane starting at $S=0$. Note that the curve continuously changes its slope when passing through the cusps.

The bottom panel of Fig. \ref{fig:U} shows the curve $(U(T,h),S(T,h))$ $N=20$, $N_S=2$ and
$h=J/2$. Only one cusp appears here in correspondence to the single extremum (a minimum) of the graph $S(T,h)$, compare to Fig. \ref{fig:S}.

In presence of vanishingly weak coupling, the specific heat is positive, meaning that the entropy is a strictly increasing function of $T$. This precludes the possibility of having local extrema in the entropy, which in turn excludes the appearance of cusps in the $S,U$ plot. 

\section{The system reduced density matrix}
\label{sec:rho}
In this section we study the density matrix of the system of interest $\mathcal{S}$. We recall that a super-bath thermalizes the total system $\mathcal{S}+\mathcal{B}$. This means that the total system is in a thermal Gibbs state:
\begin{equation}
\rho_N = e^{\beta H_N}/Z_N
\label{eq:rhoN}
\end{equation}
The density matrix $\rho$ of $\mathcal{S}$ is obtained by tracing out the bath degrees of freedom from $\rho_N$:
\begin{equation}
\rho = \Tr_{B} \rho_N \,. 
\end{equation}
For the sake of simplicity, we limit our discussion to the case of a small system with $N_S=2$, where the effects of the coupling to the bath $\mathcal{B}$ are maximal.
In this case of a system composed of two spins the reduced density matrix can be calculated by means of two-point correlators, according to the general formula
\begin{equation}
\rho = \frac{1}{4}\sum_{\alpha, \gamma= 0, x,y,z}
\langle \sigma^\alpha_1 \sigma^\gamma_2 \rangle \; \sigma^\alpha_1  \sigma^\gamma_2
\end{equation}
where $\sigma_i^0=\mathbb{1}_i$, denotes the identity operator of the Hilbert space of the $i^{\text{th}}$ spin ($i=1,2$), and $\langle \cdot \rangle$ denotes quantum expectation values with respect to $\rho_N$ given in Eq. (\ref{eq:rhoN}).
In the present case it is
\begin{equation}
\langle \sigma^\alpha_1 \sigma^\gamma_2 \rangle = \Tr \rho_N  \sigma^\alpha_1 \sigma^\gamma_2
\label{eq:bloch4}
\end{equation}
with $\Tr$ being the trace over the total system's Hilbert space.
Using Eq. (\ref{eq:bloch4}), we find the reduced density matrix (in the
basis $\{|++\rangle, |+-\rangle , |-+\rangle, |--\rangle\}$ of the common eigenvectors of $\sigma_1^z$ and $\sigma_2^z$), as:
\begin{equation}
\rho = \frac{1}{4} \left (
\begin{array}{cccc}
a_{00} & 0 & 0 & 0
\\
0 & a_{11} & a_{12}  & 0
\\
0 & a_{12} & a_{22}  & 0
\\
0 & 0 & 0  & a_{33}
\end{array}
\right)
\label{eq:rho}
\end{equation}
where
\begin{eqnarray}
a_{00} &=& 1 + \langle \sigma^z_1\rangle + \langle \sigma^z_2\rangle + \langle \sigma^z_1 \sigma^z_2\rangle \\
a_{11} &=& 1 + \langle \sigma^z_1\rangle - \langle \sigma^z_2\rangle - \langle \sigma^z_1 \sigma^z_2\rangle \\
a_{22} &=& 1 - \langle \sigma^z_1\rangle + \langle \sigma^z_2\rangle - \langle \sigma^z_1 \sigma^z_2\rangle \\
a_{33} &=& 1 - \langle \sigma^z_1\rangle - \langle \sigma^z_2\rangle + \langle \sigma^z_1 \sigma^z_2\rangle \\
a_{12} &=& 2 \langle \sigma^x_1 \sigma^x_2\rangle 
\end{eqnarray}
Using Eq. (\ref{eq:bloch4}) one  obtains, after standard but tedious algebra:
\begin{align}
\hspace*{-2cm}\langle \sigma^x_1 \sigma^x_2\rangle =&\frac{-4}{N+1} \sum_{k=1}^N 
\sin \left ( \frac{ k  \pi}{N+1} \right ) \sin\left(\frac{2 k \pi}{N+1} \right )
\mathcal{N}_k^{(N)} \label{eq:xx} \\
\langle \sigma_j^z \rangle =& -1+ \frac{4}{N+1} \sum_{k=1}^N 
\sin^2 \left (\frac{j k
  \pi}{N+1} \right ) \mathcal{N}_k^{(N)} \; \label{eq:z}
\\
\langle
\sigma^z_1 \sigma^z_2\rangle =& \langle \sigma_1^z \rangle \langle
\sigma_2^z \rangle - \langle\sigma^x_1 \sigma^x_2\rangle^2 \label{eq:zz}
\end{align}
Diagonalization of $\rho$ is straightforward. We obtain the following eigenvectors
\begin{align}
|1\rangle =& |++\rangle\\
|2\rangle =& \sin\theta_-|+-\rangle + \cos\theta_-|-+\rangle\\
|3\rangle =& \sin\theta_+|-+\rangle + \cos\theta_+|-+\rangle\\
|4\rangle =&|--\rangle
\end{align}
where the phases $\theta_\pm$ become
\begin{equation}
 \theta_\pm=\arctan \left(\frac{\langle \sigma_1^z\rangle - \langle \sigma_2^z\rangle\pm \delta }{2\langle \sigma^x_1 \sigma^x_2\rangle}\right)
\end{equation}
The corresponding eigenvalues are:
\begin{align}
p_1 =& a_{00} \\
p_2 =&  \left(1 - \delta - \langle \sigma^z_1 \sigma^z_2\rangle\right)/4\\
p_3 =&  \left(1 + \delta - \langle \sigma^z_1 \sigma^z_2\rangle\right)/4\\
p_4 =& a_{33}\\
\end{align}
where we have introduced the abbreviation:
\begin{align}
\delta = \sqrt{4\langle \sigma^x_1 \sigma^x_2\rangle^2+(\langle \sigma_1^z\rangle - \langle \sigma_2^z\rangle)^2}\\
\end{align}
\subsection{The zero temperature limit}
It is interesting to study the spectrum of the reduced density matrix at zero temperature, i.e., for $\beta \rightarrow \infty$.
\begin{figure}[t]
\includegraphics[width=8cm]{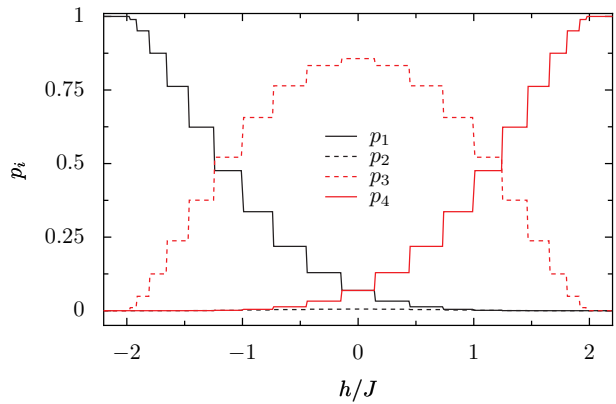}
\caption{(Color online) The low temperature spectrum of the system reduced density matrix  consists of eigenvalues that change step-like as a function of applied field $h/J$. Parameters are: $J \beta= 2000,N_S=2,N_B=20$. Beyond $|h/J|=2$ only a single nonvanishing eigenvalue exists. At higher temperatures the steps are washed out.}
\label{fig:spectrum}
\end{figure}
Fig. \ref{fig:spectrum} shows the spectrum as a function of the applied
magnetic field for $\beta = 1000$, $J=1$, and $N=20$. For $h>2J$ the
only nonzero eigenvalue is $p_4$, meaning that the open system $\mathcal S$ is in
the pure state $|4\rangle=|--\rangle$. The same holds for $h<-2J$ too, in that case the ground state is $|1\rangle=|++\rangle$. This is an interesting results:
even though the system is (strongly) coupled to the bath
$\mathcal{B}$, its ground
state is the same pure state $|--\rangle$ (or $|++\rangle$) that the system would have
in absence of the bath, if the magnetic field is strong enough. 
The fact that the bath does not alter the
ground state for $|h|>2J$ is another interesting aspect of a system strongly coupled 
to its environment.
This happens because for $|h|>2J$ the total system ground state is a pure factorized state given by the product of single spin states all pointing in the same direction (parallel to $h$ for $h<-2J$, antiparallel to $h$ for $h>2J$). By tracing out the bath spins, a pure factorized state given by aligned spins remains for the subsystem $\mathcal{S}$. On the other hand, when $|h|<2J$, the spins in the ground
state of the total system are entangled. Thus tracing out the bath degrees of freedom leads now to a mixed state for the subsystem $\mathcal{S}$.
Note that the transition of the system density matrix takes place at the same parameter value $h/(2J)=1$ at which the infinite chain ($N=\infty$) undergoes a \emph{quantum phase transition} \cite{SondhiRMP97}.

The abrupt steps appearing in the spectrum, as displayed in Fig. \ref{fig:spectrum}, correspond to the values of $h$ for which the energy of a fermionic eigenmode vanishes, i.e, $\lambda_k^{(N)}=h-2J\cos(k\pi/(N+1))=0$, see Eq. (\ref{eq:lambda}). 
These steps stem from the terms $\mathcal{N}_k^{(N)}$ in Eqs. (\ref{eq:xx},\ref{eq:z}), which
at zero temperature tend to unit steps:
\begin{equation}
 \lim_{\beta \rightarrow \infty} \mathcal{N}_k^{(N)} = \lim_{\beta \rightarrow \infty}1/({1+e^{\beta\lambda_k^{(N)}}})= \theta(-\lambda_k^{(N)})
\end{equation}
where $\theta$ is the Heaviside function.
The summation over all these steps, then originates the staircase structure of the spectrum.
In the thermodynamic limit $N_B \rightarrow \infty$, both width and height of the steps shrink and a continuous curve results. An analytical expression for the $p_i$'s can then easily be found by replacing sums with integrals (not reported).

\subsection{The high temperature limit}
It is apparent that, in general, the reduced density matrix is not of the form $\rho_0=e^{-\beta H_S}/(\Tr_S e^{-\beta H_S})$, corresponding to the uncoupled case. In the same basis $\{|++\rangle, |+-\rangle , |-+\rangle, |--\rangle\}$ of Eq. (\ref{eq:rho}), this canonical $\rho_0$ reads:
\begin{equation}
\rho_0 = \frac{1}{Q_0} \left (
\begin{array}{cccc}
e^{-\beta h} & 0 & 0 & 0
\\
0 & \cosh(\beta J) & -\sinh(\beta J)  & 0
\\
0 & -\sinh(\beta J) & \cosh(\beta J)  & 0
\\
0 & 0 & 0  & e^{\beta h}
\end{array}
\right)
\label{eq:rho0}
\end{equation}
with $Q_0= \Tr_S e^{-\beta H_S}=2[\cosh(\beta J)+ \cosh(\beta h)]$.
The analysis carried out in the previous section shows that, at high temperature, the thermodynamic behavior of the system is not affected by the coupling to the bath.
This suggests that at high temperature the reduced density matrix $\rho$ should tend to the uncoupled system density matrix $\rho_0$. 
This is indeed the case. Taylor expansion around $\beta=0$ reveals that $\rho$ and $\rho_0$ coincide up to second order in $\beta$.


\section{Conclusions}
\label{sec:conclusions}
The (possible) appearance of thermodynamic anomalies due to a violation of the the usual \emph{weak coupling} assumption have been highlighted.
Negative specific heats and entropies were reported already in the literature \cite{HIT_NJP08,IHT_PRE09,CTH_JPHYSA09}, here, for the first time, we described anomalies of the susceptibility. Apart from the fact that the susceptibility may become negative, its value can also be enhanced by several orders of magnitude, in the strong coupling regime.

The coupling strength of the XY model is determined by the exchange energy $J$. This energy has to be compared with thermal energy and the bulk energy of the system, which grows with the system size $N_S$. Therefore, the anomalies may only emerge for relatively small systems at sufficiently low tempratures. For  large systems and at high temperatures,  the canonical density matrix $\exp[-H_S/k_BT]/Z_S$
describes the equilibrium properties of the system excluding the presence of any anomaly. This is a well known generic feature of any spatially extended system interacting with its environment by short range interactions \cite{Landau5}.

The analysis of the reduced density matrix shows that the thermal equilibrium properties of open quantum systems may grossly differ from those resulting from a canonical Gibbs state.
This is true even in the thermodynamic limit of the bath $N_ B \rightarrow \infty$, as long as the coupling remains strong, ($J/k_BT$ is sufficiently large and $N_S$ sufficiently small).
In regard to the case $N_S=2$, we found that the ground state changes from a pure to a mixed state, a situation akin to a \emph{quantum phase transition} \cite{SondhiRMP97}. This transition takes place at the same value $h=2J$ of the magnetic field at which the usual phase transition in the closed isotropic XY model, is observed.

Finally we note that for the even smaller system composed of a single spin the above described anomalies continue to exist. We have restricted ourselves to the properties of spin chains and therefore disregarded the case of an open single spin system.

\section*{Acknowledgments}
The authors thank Peter H\"anggi for his constant support, for many original ideas that he has generously shared with us and for the lively scientific atmosphere at his ``Lehrstuhl''. We wish him many active years to come.
PT would like to take this opportunity to once again express his thanks for Ref. \cite{HI_ActPhysPol06} which he was given as a birthday present and which triggered his interest and the interest of many others in the thermodynamics of small quantum systems.

M.C. and P.T gratefully acknowledge financial support by the DFG via the collaborative research center SFB-486, Project A-10, via the project no. 1517/26--2, by the German
Excellence Initiative via the {\it Nanosystems Initiative Munich}
(NIM) and by the Volkswagen Foundation (project
I/80424). D.Z. is indebted to  Jos\'e Luis
Garc\'{\i}a-Palacios for useful discussions and acknowledges
financial support from FIS2008-01240  and FIS2009-13364-C02-01 (MICINN).
 \appendix

 \section{Solution of the isotropic XY model}
 \label{App:JW}

In this appendix we briefly review the solution
of the isotropic XY model, Hamiltonian (\ref{XYmodel}) \cite{LiebShultzMattis61,mikeska}.
New operators are defined by
means of the Jordan-Wigner
transformation \cite{LiebShultzMattis61}:
\begin{equation}
 a_j = \prod _{k=1}^{j-1}\sigma^z_j \sigma_j^{-}
\end{equation}
where $\sigma_j^{-}$ is the $j^{th}$ spin lowering operator
\begin{equation}
 \sigma_j^{-}= \frac{1}{2}(\sigma^x_j+i\sigma^y_j) \, .
\end{equation}
The operators $a_j$ are fermionic operators satisfying the canonical
anti-commutation rules:
\begin{equation}
 \{a_j^\dagger,a_k\}= \delta_{i,k} \qquad \{a_j,a_k\} = 0.
\end{equation}
In terms of these operators the Hamiltonian (\ref{XYmodel}) is
expressed as:
\begin{equation}
H_N = -h \sum_{j=1}^{N} a_j^{\dagger}a_j - J
\sum_{j=1}^{N-1}(a_j^{\dagger}a_{j+1}+a_{j+1}^{\dagger}a_j)+\frac{N h}{2} \, .
\end{equation}
Next, following known procedures, yet new operators are defined, via the discrete
sine Fourier transform:
\begin{equation}
 b_k= \sqrt{\frac{2}{N+1}} \sum_{i=1}^{N} \sin \left( \frac{ki\pi}{N+1}
\right)a_i \, .
\end{equation}
The new operators also obey canonical fermionic anti-commutation rules:
\begin{equation}
 \{b_j^\dagger,b_k\}= \delta_{i,k} \qquad \{b_j,b_k\} = 0.
\end{equation}
Expressing the Hamiltonian in terms of the $b_k$'s, one gets:
\begin{equation}
H_N =  \sum_{j=1}^{N}\lambda_k^{(N)} 
b_j^{\dagger}b_j - \frac{N h}{2} \, ,
\label{eq:HN3}
\end{equation}
{\it i.e.} a free fermionic Hamiltonian, with single mode energies $\lambda_k^{(N)}$, Eq. (\ref{eq:lambda}). 
Denoting the Fock state associated to the operators
$\{b_k\}_{k=1..N}$ as $|n_1, \dots n_N \rangle$,
that is:
\begin{equation}
b_k^\dagger b_k |n_1, \dots n_k \dots n_N \rangle= n_k |n_1, \dots n_k \dots n_N \rangle
\end{equation}
with $n_k=0,1$, the eigenvalues of the Hamiltonian read:
\begin{equation}
\varepsilon_{n_1, \dots n_N} = \sum_{j=1}^{N}\lambda_k^{(N)}  n_j + \frac{N h}{2}\, .
\end{equation}

\end{document}